\documentclass[11pt]{article}

\usepackage[T1]{fontenc}
\usepackage[utf8]{inputenc}
\usepackage{lmodern}
\usepackage{microtype}
\usepackage{textcomp}

\usepackage{amsmath,amssymb,amsfonts,amsthm}

\usepackage[letterpaper,margin=1in]{geometry}

\usepackage{xcolor}
\definecolor{linkblue}{HTML}{1A4F9C}
\definecolor{calloutrule}{HTML}{4A4A4A}
\definecolor{calloutbg}{HTML}{F4F3F1}

\usepackage[colorlinks=true,
            linkcolor=linkblue,
            citecolor=linkblue,
            urlcolor=linkblue,
            breaklinks=true,
            pdfborder={0 0 0}]{hyperref}
\usepackage{url}

\usepackage{graphicx}
\graphicspath{{figures/}{./}}
\setkeys{Gin}{keepaspectratio}

\usepackage{caption}
\usepackage{booktabs}
\usepackage{longtable}
\usepackage{array}
\usepackage{calc}
\usepackage{enumitem}

\usepackage{mdframed}
\usepackage{etoolbox}   

\newenvironment{theorem}[1][]%
  {\ifstrempty{#1}{}{\noindent\textbf{#1.}\space}}{}
  {\ifstrempty{#1}{}{\noindent\textbf{#1.}\space}}{}

\surroundwithmdframed[%
  topline=false, bottomline=false, rightline=false, leftline=true,
  linewidth=2pt, linecolor=calloutrule,
  backgroundcolor=calloutbg,
  skipabove=1em, skipbelow=1em,
  innerleftmargin=12pt, innerrightmargin=10pt,
  innertopmargin=8pt, innerbottommargin=8pt
]{definition}
\surroundwithmdframed[%
  topline=false, bottomline=false, rightline=false, leftline=true,
  linewidth=2pt, linecolor=calloutrule,
  backgroundcolor=calloutbg,
  skipabove=1em, skipbelow=1em,
  innerleftmargin=12pt, innerrightmargin=10pt,
  innertopmargin=8pt, innerbottommargin=8pt
]{theorem}

\usepackage{titlesec}
\titleformat{\section}{\normalfont\large\bfseries}{\thesection}{0.8em}{}
\titleformat{\subsection}{\normalfont\normalsize\bfseries}{\thesubsection}{0.8em}{}
\titleformat{\subsubsection}{\normalfont\normalsize\itshape}{\thesubsubsection}{0.8em}{}
\titlespacing*{\section}{0pt}{1.6em}{0.5em}
\titlespacing*{\subsection}{0pt}{1.2em}{0.4em}

\usepackage{fancyhdr}
\fancypagestyle{plain}{%
  \fancyhf{}%
  \fancyfoot[C]{\footnotesize\thepage}%
}

\NewDocumentCommand\citeproctext{}{}
\NewDocumentCommand\citeproc{mm}{%
  \begingroup\def\citeproctext{#2}\cite{#1}\endgroup}
\makeatletter
 \let\@cite@ofmt\@firstofone
 \def\@biblabel#1{}
 \def\@cite#1#2{{#1\if@tempswa , #2\fi}}
\makeatother
\newlength{\cslhangindent}
\newlength{\csllabelwidth}
\newenvironment{CSLReferences}[2] 
 {\begin{list}{}{%
  \setlength{\itemindent}{0pt}
  \setlength{\leftmargin}{0pt}
  \setlength{\parsep}{0pt}
  \ifodd #1
   \setlength{\leftmargin}{\cslhangindent}
   \setlength{\itemindent}{-1\cslhangindent}
  \fi
  \setlength{\itemsep}{#2\baselineskip}}}
 {\end{list}}

\hypersetup{
  pdftitle={Bits per Spike as a Betting Game},
  pdfauthor={Alex H. Williams},
  pdfkeywords={bits per spike, held-out log-likelihood, e-values,
anytime-valid inference, game-theoretic statistics, generalized linear
models, spike trains},
  pdfcreator={},
  pdfproducer={}
}

\title{\vspace{-2em}Bits per Spike as a Betting
Game\\[0.45em]{\large\normalfont\itshape An interpretable unit for
held-out log-likelihood in neural data analysis}\vspace{-0.4em}}
\author{Alex H. Williams\\[0.3em]\normalsize Center for Neural Science,
New York University\\[0.3em]\normalsize Center for Computational
Neuroscience, Flatiron
Institute\\[0.2em]\normalsize\texttt{alex.h.williams@nyu.edu}}
\date{July 29, 2026}

\begin{document}

\maketitle
\thispagestyle{plain}

\begin{abstract}
\noindent Held-out log-likelihood is the standard currency for comparing
statistical models of neural spike trains, and is often reported as
\emph{bits per spike} relative to a homogeneous Poisson baseline. The
units of this metric are difficult to reason about: it is rarely obvious
whether an improvement of, say, \(0.34\) bits per spike is a large
effect or a negligible one. This note develops an interpretation of
held-out log-likelihood borrowed from game-theoretic statistics. If
\(\mathcal{L}\) denotes a model's expected log-likelihood ratio to the
homogeneous Poisson baseline, then we show that
\(-\log(\alpha) / \mathcal{L}\) represents the number of held-out time
bins of recording needed to reject the baseline at level \(\alpha\)
under a particular null hypothesis testing procedure based on a betting
game. Thus, a simple re-scaling of the heldout log-likelihood yields an
intuitive metric of model performance in units of recording time which
answers ``how much heldout data would I need, on average, to disprove
the null.'\,' We illustrate the construction on head-direction cells
recorded in mouse anterior thalamus, where a generalized linear model
reaches significance in roughly \(120\) ms of held-out data for a
strongly tuned cell and roughly \(11\) s for a moderately tuned cell.
\end{abstract}

\begingroup
\par
\endgroup
\vspace{1em}

\begingroup
\small
\noindent\rule{\linewidth}{0.4pt}\par
\vspace{0.4em}
\noindent This material is also published as a blog post, \emph{Bits per
Spike as a Betting Game}, on
\url{https://neurostatsblog.github.io/2026/05/27/model-comparison-by-betting/}.\par
\vspace{0.3em}
\noindent\rule{\linewidth}{0.4pt}\par
\endgroup
\vspace{1.2em}

Whenever we fit a model to neural or behavioral data, we need to
benchmark it against simpler or well-known baselines. Typically this is
done by reporting the difference in log-likelihoods on heldout data. For
example, the popular ``bits per spike'' performance metric (see e.g.
\citeproc{ref-pillow2008}{Pillow et al. 2008}) is simply the log (base
2) likelihood of the model minus the log (base 2) likelihood of a
homogeneous Poisson process (or another appropriate baseline model),
divided by the total number of spikes in the dataset.

This note offers some thoughts on how we can interpret this performance
measure. For example, if my model gives a 0.34 bits per spike
improvement over the baseline, should I interpret that as very good?
Marginal? Completely inconsequential?

This note will focus on a particular interpretation that imagines the
model playing a betting game against the ``market'' defined by the
baseline model. This game-theoretic framing has gained traction in a
certain corner of statistics (see the references for a brief list of
introductory resources).

\section{Basic Setup}\label{basic-setup}

Suppose that we have fit a model \(Q\) and a baseline \(B\) on training
data and that we're now ready to compare them head-to-head on heldout
test data. Let \(X_1, X_2, X_3, \dots\) denote a (potentially infinite)
sequence of heldout data samples. We assume these samples are
independent and identically distributed according to an unknown
distribution \(P\). Our hope is that \(Q\) is in some sense ``closer''
to the true distribution \(P\) than the baseline model \(B\).

For simplicity, we'll assume that \(Q\) and \(B\) have strictly positive
probability density or mass functions \(q(x)>0\) and \(b(x)>0\). This
allows us to compute \emph{likelihood ratios}, \(q(X)/b(X)\) for
\(X \sim P\), without having to worry about divide-by-zero
errors.\footnote{It is possible to make the math work without making
  this assumption. Generally, this would let us turn ``\(>\) relations''
  into ``\(\geq\) relations'' and likewise turn ``\(\leq\) relations''
  into ``\(<\) relations''. However, for simplicity we just stick to
  strict inequalities for the purpose of this note.}

A nice measure of performance is the expected log-likelihood ratio:
\begin{equation}
\mathcal{L} = \mathbb{E}_{X \sim P} \Big [ \log q(X)/ b(X) \Big ] = \mathbb{E}_{X \sim P} \Big [ \log q(X) - \log b(X) \Big ] 
\label{eq:expected-log-likelihood-ratio}
\end{equation} Note that the expectation is computed under \(P\). Since
\(P\) is unknown in real world situations, we can estimate the
expression above by holding out a test set with \(T\) data samples and
approximating the expectation with an empirical average:
\begin{equation}
\mathcal{L} \approx \widehat{\mathcal{L}} = \frac{1}{T} \sum_{t=1}^T \log q(X_t)/ b(X_t)  = \frac{1}{T} \sum_{t=1}^T \Big [ \log q(X_t) - \log b(X_t) \Big ] 
\label{eq:empirical-log-likelihood-ratio}
\end{equation}

We have thus far used natural logarithms, but substituting base-2
logarithms can aid interpretation. By the change of base formula,
\(\mathcal{L}_2 = \mathcal{L} / \log(2)\) is the expected base-2 log
likelihood. The base-2 log likelihood has units of ``bits'', and
normalizing by the number of spikes gives the popular ``bits per spike''
metric.

Even more intriguingly, we will be able to draw a connection to null
hypothesis testing at significance threshold \(\alpha\) (for example,
\(\alpha = 0.05\)). It turns out that the reciprocal of the expected
base-\((1/\alpha)\) log likelihood,
\(\mathcal{L}_{1/\alpha}^{-1} = -\log(\alpha) / \mathcal{L}\), will have
a very satisfying interpretation as the number of heldout data samples
needed to reject the baseline model as a null hypothesis.

\emph{You don't need to understand the betting game to use this result!}
If you wish to skip the derivation, you can jump straight to the
\hyperref[take-home-message]{\textbf{Take Home Message}} at the end of
this note.

\section{A Demo of the Betting Game}\label{a-demo-of-the-betting-game}

Our goal is to come up with intuitive interpretations of \(\mathcal{L}\)
as a measure of model performance. One way to approach this is to
imagine model \(Q\) as a ``player'' in a betting game. On each round of
the game, we sample a heldout datapoint \(X \sim P\) and pay the player
based on how well they predicted the outcome. Each betting contract
comes at a price, which is determined by the baseline model \(B\).
Intuitively, the more that \(Q\) is able to ``beat the market'' by
accumulating wealth, the stronger the evidence we have in favor of \(Q\)
over \(B\).

Before we get to the precise details of the game, let's look at a few
examples. Figure 1 shows tuning curves fit to three example
head-direction cells from the mouse anterior thalamus
(\citeproc{ref-peyrache2015}{Peyrache et al. 2015}), distributed through
the \textbf{nemos} (\citeproc{ref-nemos}{NeMoS Developers 2025})
library. Each cell's spiking is modulated by the animal's current head
direction, and binning spike counts by head-direction angle gives
\emph{empirical firing rates} (grey dots). We then discretized spike
trains using \(\Delta = 0.05\) second time bins and used
\href{https://nemos.readthedocs.io/}{\textbf{nemos}} to fit a GLM model
with cyclic B-spline basis functions (solid line, \(Q\)), and a
homogeneous Poisson process as a baseline model (dashed grey line).

Concretely, model \(Q\) is a Poisson GLM whose single covariate is the
animal's head direction in each bin, expanded in ten cyclic B-spline
basis functions and fit by unregularized maximum likelihood with L-BFGS.
The baseline \(B\) is a homogeneous Poisson process whose rate is the
mean spike count per bin, estimated on the same training half. Each
cell's recording was split 50/50 into training and test halves after
shuffling the order of time bins, and the whole procedure was repeated
for ten independent random splits. The three cells were selected by
their degree of head-direction modulation: the most strongly modulated
unit, a unit at the median of the modulation distribution, and the least
modulated unit subject to a floor of 200 spikes.

\begin{figure}[t]
\centering
\includegraphics[width=\textwidth]{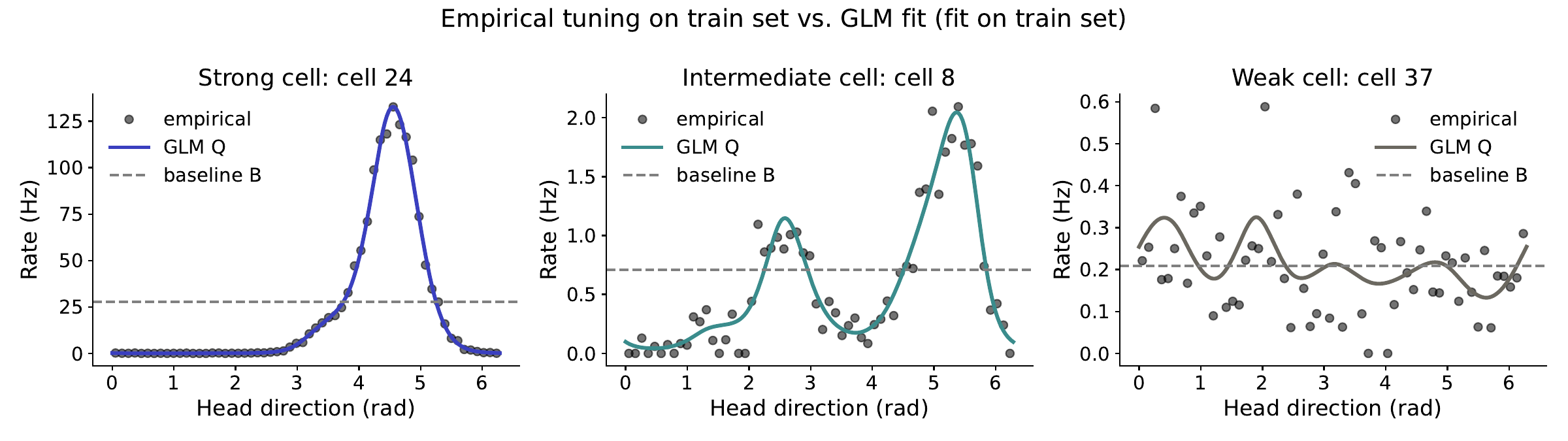}
\caption{Three example cells with strong, intermediate, and weak head direction
modulation. For each cell we fit a GLM model, \(Q\), and a flat baseline
model, \(B\).}
\end{figure}

The plots above were generated using 50\% of the observations as
``training data''. The remaining 50\% of the observations serve as
heldout ``test data'' which we use below to play the forecasting game.

In the game, player \(Q\) gambles their wealth over discrete rounds of
betting. In each round, the player forecasts and places a bet on the
number of spikes they'll see in the next time bin. Intuitively, if \(Q\)
is a much better model of the data than \(B\) then the player has an
``edge'' that they can exploit and generate returns very quickly---in
fact, it turns out to be exponentially fast.

Throughout, we will use \(W_t\) to denote the wealth of the player at
round \(t\) of the game. In Figure 2, we plot how the player's wealth
\(W_t\) evolves across the test set for each of the three cells. Every
player starts with \(W_0 = 1\) and the y-axis is on a \(\log_2\) scale,
so a value of \(\log_2 W_t = 10\) means the player has doubled their
wealth ten times (a \(2^{10} = 1024\times\) return). For each cell we
ran ten independent random train/test splits of the recording and
overlaid all ten trajectories.\footnote{It is worth pointing out an
  important approximation. We use ten independently generated train and
  test splits to generate each random wealth trajectory. Thus, we fit
  ten different GLM models and have them ``play the betting game'' on
  ten different heldout test sets. Thus, some of the variability in the
  trajectories is due to differences in the training data. Our
  exposition of the betting game does not account for this. Another
  approximation worth noting is that we do not account for
  autocorrelation in the observations---we shuffle the order of time
  bins on each split. Extending the betting game to time series data
  with autocorrelation could be important for comparing certain models
  of neural data.}

Notice that the x-axis range differs across panels --- the strong cell's
wealth game unfolds over 100 time bins (5 seconds), while the weak
cell's takes the full \textasciitilde15 minutes of test data.

\begin{figure}[t]
\centering
\includegraphics[width=\textwidth]{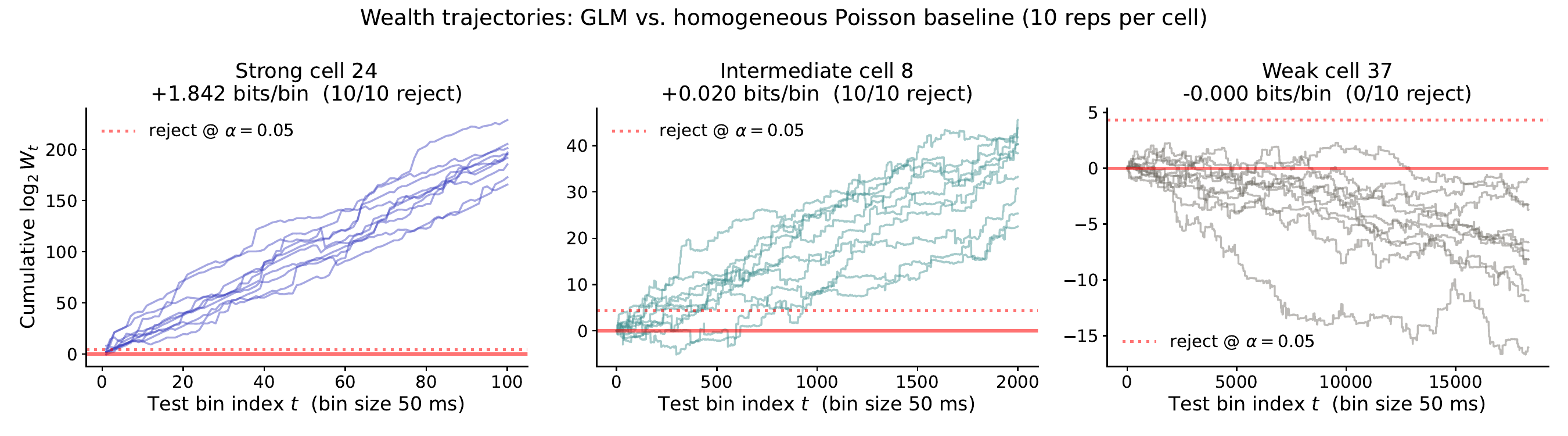}
\caption{Wealth trajectories for the three example cells, ten random train/test
splits each. The y-axis is on a \(\log_2\) scale (number of doublings of
the starting wealth). The dashed red horizontal line at
\(\log_2 W_t \approx 4.3\) marks a threshold we will motivate later.
Each panel uses its own x-axis range so the dynamics of each cell are
visible at an appropriate scale.}
\end{figure}

For the \textbf{strong cell} (left panel), the player's wealth grows
very fast---roughly linearly on the \(\log_2\) scale at a rate of
\textasciitilde{}\(1.8\) units of log wealth per round. Each round, the
player nearly quadruples their wealth (on average)! The ten trajectories
cluster tightly because the underlying signal is strong enough that
essentially any random split of the recording produces nearly the same
model and the same betting edge.

For the \textbf{intermediate cell} (middle panel) we see the same
qualitative pattern --- linear growth on the \(\log_2\) scale --- but at
a much shallower slope of about \(0.02\) units of log wealth per round.
This means that it takes the player \(1/0.020 = 50\) rounds to double
their wealth, on average.

For the \textbf{weak cell} (right panel), the player has effectively no
edge at all. Wealth wanders around the starting value of \(W_0 = 1\) and
drifts slightly \emph{downward} on average; across all ten splits, the
trajectories stay well below the dashed threshold line at the top of the
panel. This is the behavior we hope to see whenever a model offers no
genuine improvement over the baseline: betting on a useless predictor
is, in the long run, a losing strategy.

By the end of this note, we'll see that expected log likelihood ratios
describe the rate of wealth growth. For example, \(1 / \mathcal{L}_2\)
describes the number of rounds of betting needed (on average) to double
the player's wealth. Similarly, if we normalize by time bin size,
computing \(\Delta / \mathcal{L}_2\), we obtain an estimate of how much
heldout data we need (in recording duration) to double the player's
wealth.

We can also connect this to a null hypothesis test with significance
level \(\alpha\). It turns out that if the player ever generates more
than \(1/\alpha\) units of wealth, we can reject the null hypothesis
that the true distribution, \(P\), is equal to the baseline model,
\(B\), with type-I error rate less than \(\alpha\). For example, if we
set \(\alpha=0.05\), then we can reject the null hypothesis if we ever
obtain \(W_t > 1/\alpha = 20\).

In light of this, the expected base-\((1/\alpha)\) log likelihood ratio,
denoted \(\mathcal{L}_{1/\alpha}\), tells us the rate at which we
accumulate evidence to reject the null hypothesis that \(P=B\).
Similarly, \(1 / \mathcal{L}_{1/\alpha}\) and
\(\Delta / \mathcal{L}_{1/\alpha}\) respectively tell us how much data
we need (on average) to reject the null in units of time bins and
seconds. Figure 3 below re-plots the data from Figure 2 on this log
scale. Further, the x-axis is plotted in units of recorded time (by
dividing by time bin size \(\Delta\)). The units of the y-axis now
correspond to the number of times the player has accumulated enough
wealth to reject the null hypothesis at significance level \(\alpha\).
For the strong cell (\emph{left}) that timescale is around 120 ms; for
the intermediate cell it is around 11 seconds; for the weak cell the
trajectories never cross the threshold, and we therefore fail to reject
the null.

\begin{figure}[t]
\centering
\includegraphics[width=\textwidth]{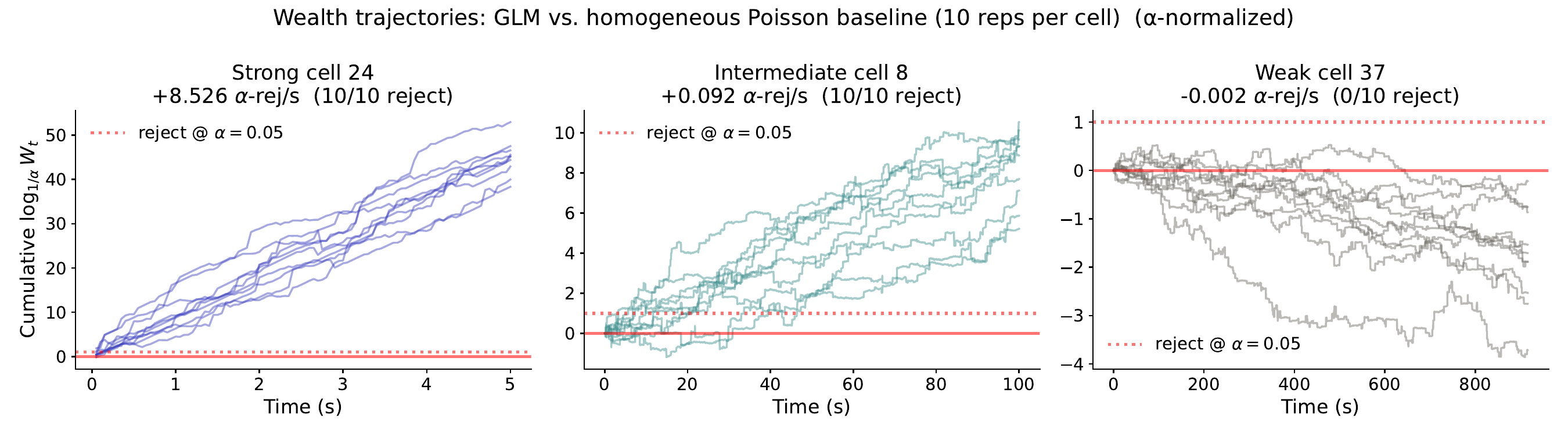}
\caption{The same wealth trajectories as Figure 2, now plotted in units of
\(\log_{1/\alpha} W_t\) with \(\alpha = 0.05\). The dashed red line at
\(y=1\) marks the wealth required to reject the null at level
\(\alpha\). The slope of each trajectory is the rate at which the player
accumulates rejection-worth of evidence.}
\end{figure}

A common alternative normalization is \emph{bits per spike}, the metric
mentioned at the top of the post. If \(\bar\lambda_b\) is the baseline's
expected spike count per bin, then
\(\text{bits/spike} = \mathcal{L}_2 / \bar\lambda_b\) coincides with the
player's wealth growth per spike observed, rather than per bin elapsed.
Normalizing by the number of spikes is potentially useful when comparing
models fit to neurons with very different firing rates. However, since
the betting game is naturally indexed by time bins of the recording, the
rest of the post will stick to normalizing the log likelihood ratio in
units of time.

Now that we've sketched some results, it's time to make this picture
rigorous. Specifically, we want to (1) define exactly how each round of
betting works, (2) explain the optimal strategy that player \(Q\) can
implement, and (3) understand what the significance threshold line in
the figure means in terms of a formal hypothesis test.

\section{Introducing the Game}\label{introducing-the-game}

The game starts by giving player \(Q\) one unit of wealth: \[
W_0 = 1
\] At each round of the game, player \(Q\) uses all of their wealth to
purchase \emph{prediction contracts}, specified by a function
\(C(x) > 0\). On each round, indexed by \(t\), we sample an observation
\(X_t \sim P\) and pay the player \(C(X_t)\) units of wealth. Thus, the
player receives a random sequence of returns
\(C(X_1), C(X_2), C(X_3), \dots\) over discrete rounds of the game. The
wealth updates according to: \begin{align}
W_t &= \big ( \, W_{t-1} / \pi(C) \, \big ) \cdot C( X_{t} ) . \label{eq:wealth-process-verbose}
\end{align} where \(\pi(C)\) denotes the \emph{price} of the contract.

Equation \eqref{eq:wealth-process-verbose} is simple. The first term,
\(W_{t-1} / \pi(C)\), is the \emph{number of contracts} the player is
able to purchase given their previous wealth at round \(t - 1\). The
second term, \(C(X_t)\), is the payoff of each contract. Note that we
allow the wealth and the number of purchased contracts to be infinitely
divisible into fractions.

Intuitively, the price of a contract is set by what people are willing
to buy and sell it for, which reflects their expectations about the
underlying distribution \(P\). For the purposes of our game we'll assume
that the market consensus---or the ``wisdom of the crowd''---coincides
with the baseline model \(B\). Formally, it turns out that the fair
price of a contract is given by its expected value under the market
consensus distribution. That is, \begin{equation}
\pi(C) = \mathbb{E}_{X \sim B} \, \big [ \,  C(X) \, \big ] . \label{eq:general-pricing-constraint}
\end{equation} See \hyperref[supplementary-note-1]{\textbf{Supplementary
Note 1}} for a quick derivation of this equation.

The pricing constraint in equation \eqref{eq:general-pricing-constraint}
allows us to simplify the structure of the game by assuming that the
contracts have unit price. Indeed, for any contract \(C(\cdot)\), we can
define a new contract \(S(x) = C(x)/\pi(C)\) which has unit price,
\(\pi(S) = 1\). The wealth update for \(C\) and \(S\) is equivalent
since \[
\big ( \, W_{t-1} / \pi(C) \, \big ) \cdot C( X_{t} ) = W_{t-1} \cdot S(X_t),
\] by the definition of \(S\). Therefore, for the rest of this note we
will focus on the simplified wealth process

\begin{theorem}[]
\textbf{Simplified wealth process.} Assuming that the player purchases
positive contracts \(S(x) > 0\) of unit price, i.e. \begin{align}
\mathbb{E}_{X \sim B} \big [ \, S(X) \, \big ] = 1 \label{eq:unit-price-constraint}
\end{align} then the player's wealth evolves according to \begin{align}
W_t &=  W_{t-1} \cdot S( X_{t} ) . \label{eq:wealth-process}
\end{align}
\end{theorem}

\section{Choosing the optimal contract
function}\label{choosing-the-optimal-contract-function}

Player \(Q\) is allowed to choose the function \(S(\cdot)\) however they
like, so long as it satisfies \eqref{eq:unit-price-constraint}. Out of
this space of feasible contracts, which one should \(Q\) choose to play?

After \(T\) rounds of betting according to equation
\eqref{eq:wealth-process}, the player will accumulate \begin{align}
W_T = \prod_{t=1}^T S( X_{t} ) 
      &= \exp \log \prod_{t=1}^T S( X_{t} ) \\
      &= \exp \sum_{t=1}^T \log S( X_{t} ) \\
      &= \exp \Big ( T \cdot \Big ( \tfrac{1}{T} \sum_{t=1}^T \log S(X_t) \Big ) \Big ) \\
      &\approx \exp \Big ( T \cdot \mathbb{E}_{X \sim P} \log S(X) \Big )
      \label{eq:q-wealth-growth}
\end{align} units of wealth. The approximation in the final line comes
from replacing the empirical expectation
\(\tfrac{1}{T} \sum_{t=1}^T \log S(X_t)\) with the true expected value
under \(P\).

Since the player believes that \(P = Q\), they anticipate that their
wealth can grow exponentially over time according to: \begin{equation}
W_T \approx \exp \Big ( T \cdot \mathbb{E}_{X \sim Q} \log S(X) \Big )
\end{equation} To maximize their rate of wealth growth, a reasonable
strategy is to choose \(S(\cdot)\) in order to \begin{align}
\text{maximize} ~~ \mathbb{E}_{X \sim Q} \log S(X) \quad \text{subject to } \eqref{eq:unit-price-constraint} 
\label{eq:kelly-criterion}
\end{align}

Quite pleasingly, as shown in
\hyperref[supplementary-note-2]{\textbf{Supplementary Note 2}}, the
solution to this optimization problem, denoted \(S^\star\), turns out to
be the likelihood ratio! \begin{equation}
S^\star(x) = \frac{q(x)}{b(x)}
\label{eq:betting-function-equals-likelihood-ratio}
\end{equation} Combining equations
\eqref{eq:betting-function-equals-likelihood-ratio} and
\eqref{eq:q-wealth-growth} with the definition of \(\mathcal{L}\) in
\eqref{eq:expected-log-likelihood-ratio}, we see that the long-run
wealth of the player is approximated by \begin{equation}
W_T \approx \exp \Big ( \mathcal{L} \cdot T \Big )
\label{eq:player-q-long-term-wealth}
\end{equation} for large \(T\). In other words, the wealth accumulated
by player \(Q\) in the game will, over the long run, grow or decay
exponentially fast at a rate given by the expected log-likelihood ratio.

\section{Some interpretations}\label{some-interpretations}

We are now in a position to rigorously reinterpret Figure 2. Taking
logarithms on both sides of \eqref{eq:player-q-long-term-wealth} and
changing to base-2 yields a linear relationship between the log-wealth
and time bin index: \begin{equation}
\log_2 ( W_T ) \approx \mathcal{L}_2 \cdot T .
\end{equation} The reciprocal \(1/\mathcal{L}_2\) is therefore the
average \emph{doubling time} of the wealth process in units of time bins
(or equivalently, rounds of betting). Multiplying by the time bin size,
\(\Delta/\mathcal{L}_2\), changes the units of the doubling time to
seconds of recording. For example, in Figure 2, the strong cell
(\emph{left}) doubles its wealth every \textasciitilde{}\(25\) ms while
the intermediate cell (\emph{middle}) doubles its wealth every
\textasciitilde{}\(2.5\) seconds.

It is also interesting to note a connection to the KL divergence. While
\(\mathcal{L}\) determines the \emph{actual} rate of wealth growth, the
player's anticipated rate of wealth growth is given by replacing \(P\)
in \eqref{eq:q-wealth-growth} with \(Q\), yielding: \begin{equation}
\mathbb{E}_{X \sim Q} \Big [ \log q(X)/ b(X) \Big ]
\end{equation} which is precisely the KL divergence,
\(D_{\mathrm{KL}}(Q \,\Vert\, B)\).\footnote{This interpretation of KL
  divergence is due to JL Kelly Jr.~in a 1956 tech report
  (\citeproc{ref-kelly1956}{Kelly 1956}). The principle that the player
  should choose their bet to maximize the term in the exponent appearing
  in \eqref{eq:q-wealth-growth} is named after him---it is known as the
  Kelly criterion in quantitative finance.}

Figure 4 shows the same wealth trajectories as in Figure 3 with the
expected rate of wealth growth, given by
\(D_{\mathrm{KL}}(Q \,\Vert\, B)\), overlaid as a dark black line. For
the strong and intermediate cells (\emph{left} and \emph{middle}
panels), the anticipated and realized growth rates agree, indicating
that the GLM model is well-calibrated on this dataset. For the weak
cell, the anticipated line climbs gradually above zero while the
realized trajectories drift slightly below. This is suggestive of
overfitting: \(Q\) believes it has a tiny edge that doesn't actually
exist on heldout data.

\begin{figure}[t]
\centering
\includegraphics[width=\textwidth]{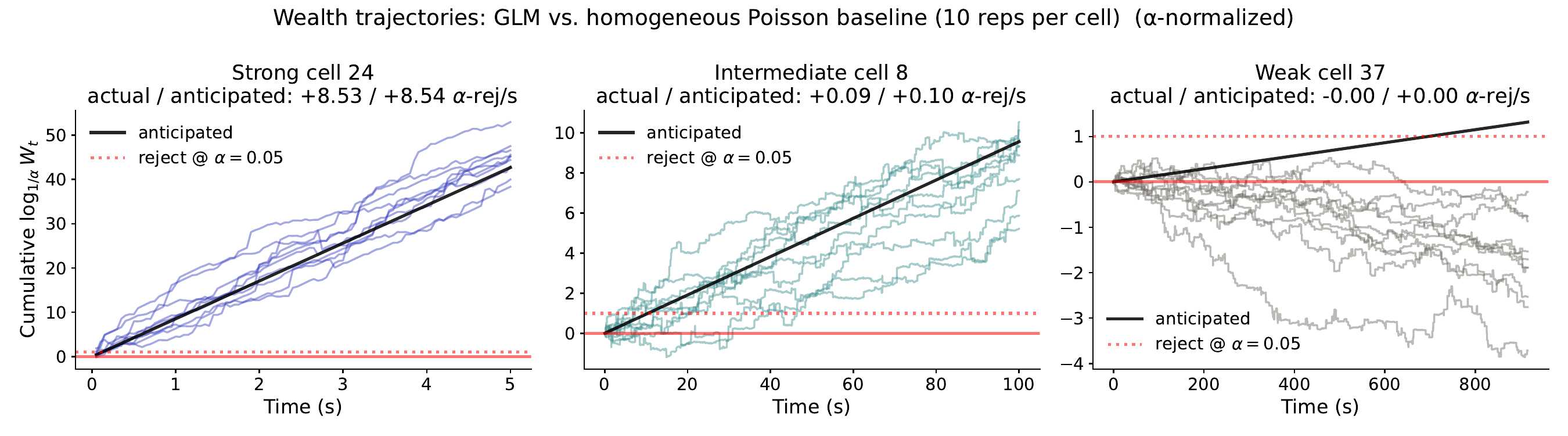}
\caption{Same wealth trajectories as Figure 3, now with the player's anticipated
rate of wealth growth overlaid as a black line. The slope of the line is
\(D_{\mathrm{KL}}(Q \,\Vert\, B)\) per round of betting, averaged across
the ten train/test splits. For the strong and intermediate cells, the
anticipated and realized trajectories agree closely. For the weak cell,
the anticipated line drifts slightly above zero while the realized
trajectories drift slightly below --- a small but real signature of
overfitting on the training data.}
\end{figure}

\section{Connection to Hypothesis
Testing}\label{connection-to-hypothesis-testing}

We have not yet drawn a rigorous connection to null hypothesis testing,
as promised by the plots in Figures 3 and 4. The connection is simple
and follows concretely from a result called Ville's inequality
(\citeproc{ref-ville1939}{Ville 1939}; \citeproc{ref-howard2020}{Howard
et al. 2020}), summarized below.

\begin{theorem}[]
\textbf{Ville's Inequality (informal).} Let \((M_t)_{t \geq 0}\) be a
sequence of random variables such that \(M_t \geq 0\) almost surely for
all \(t\) and
\(\mathbb{E}[M_{t} \mid M_0, \dots, M_{t-1}] \leq M_{t-1}\) for all
\(t\). Then for any \(\alpha > 0\), \[
\textrm{Pr}\!\left( \sup_{t \geq 0} M_t \, \geq \, 1/\alpha \right) \, \leq \, \alpha \cdot \mathbb{E}[M_0] .
\]
\end{theorem}

Intuitively, this result states that if each step of the random sequence
\((M_t)_{t \geq 0}\) is not increasing in expectation,\footnote{For
  those who appreciate jargon, we call \((M_t)_{t \geq 0}\) a
  \emph{supermartingale} if it satisfies
  \(\mathbb{E}[M_{t} \mid M_0, \dots, M_{t-1}] \leq M_{t-1}\). In the
  stricter case where the inequality is saturated,
  i.e.~\(\mathbb{E}[M_{t} \mid M_0, \dots, M_{t-1}] = M_{t-1}\), we call
  \((M_t)_{t \geq 0}\) a \emph{martingale}. Ville's inequality says that
  all nonnegative supermartingales (and martingales) are upper bounded
  for all time with high probability.} then the probability of
\((M_t)_{t \geq 0}\) \emph{ever} crossing above a high threshold is
small, even if we run the simulation infinitely far into the future.
\hyperref[supplementary-note-3]{\textbf{Supplementary Note 3}} sketches
a quick proof of Ville's inequality.

To see why this matters in our setting, suppose that the baseline \(B\)
is in fact the true data-generating distribution---i.e.~\(P = B\). Then
we would have \[
\mathbb{E}_{X_t \sim B} \big[ W_t \mid W_0 \dots W_{t-1} \big]
\,=\, W_{t-1} \cdot \mathbb{E}_{X_t \sim B} \left[ \frac{q(X_t)}{b(X_t)} \right]
\,=\, W_{t-1} .
\] since the expectation of the likelihood ratio equals one.\footnote{Concretely,
  \(\mathbb{E}_{X \sim B} \left[ \frac{q(X)}{b(X)} \right] = \int \frac{q(x)b(x)}{b(x)} dx =  \int q(x) dx = 1\).}
This means that we can apply Ville's inequality, which tells us that the
probability of player \(Q\)'s wealth \emph{ever} exceeding the threshold
\(1/\alpha\)---at any round of the game, even infinitely far into the
future---is at most \(\alpha\): \[
P\!\left( \sup_{t \geq 0} W_t \, \geq \, 1/\alpha \right) \, \leq \, \alpha .
\] This furnishes an \emph{anytime-valid} hypothesis test
(\citeproc{ref-ramdas2023}{Ramdas et al. 2023}) of the null
\(H_0 : P = B\). We may reject \(H_0\) at level \(\alpha\) as soon as
\(W_t\) crosses \(1/\alpha\), regardless of how many rounds have been
played. Unlike classical fixed-sample tests, we are free to peek at the
data, stop early, or keep collecting more samples, all without inflating
the type I error rate. For example, if we use \(\alpha = 0.05\) (as is
customary), then we can reject the null hypothesis that \(P = B\) if
player \(Q\)'s wealth \emph{ever} exceeds 20.

\section{Take Home Message}\label{take-home-message}

A central message of this note is that the expected log-likelihood
ratio, \(\mathcal{L}\), can be interpreted as the exponential rate at
which a player generates wealth in a betting game based on forecasting
future events. For example, \(\log(2)/\mathcal{L}\) defines the time it
takes for the player to double their wealth (in units of time bins).

The betting game interpretation is satisfying, but it involves some
effort to conceptualize and formally derive. In an effort to simplify
the take home message, I propose two main summary statistics: the
\textbf{samples to significance}, \(\tau\), and the \textbf{time to
significance}, given by \(\tau \, \Delta\).

\begin{theorem}[]
\textbf{Time to Significance.} Let \(\mathcal{L} > 0\) denote the
expected log-likelihood ratio of a model \(Q\) relative to baseline
\(B\), as in equation \eqref{eq:expected-log-likelihood-ratio}. The
\emph{samples to significance}, \[
\tau = \frac{-\log(\alpha)}{\mathcal{L}} ,
\] reflects the number of heldout samples needed on average to reject
the null hypothesis \(H_0 : P = B\) at significance level \(\alpha\). If
each sample is a time bin of \(\Delta\) seconds, the corresponding
\emph{time to significance} is given by
\(\tau \, \Delta = -\Delta \log(\alpha) / \mathcal{L}\).
\end{theorem}

In practice, we estimate these quanitites by first fitting \(Q\) and
\(B\) on training data and using heldout test data to compute an
estimate of the expected log-likelihood ratio, \(\widehat{\mathcal{L}}\)
in \eqref{eq:empirical-log-likelihood-ratio}. The time to significance
is a summary statistic of model performance that can be used in place
of, or complementary to, the conventional ``bits per spike'' metric.
This gives \(\tau \, \Delta \approx 120\) ms for the strong example cell
and \(\tau \, \Delta \approx 11\) s for the intermediate example cell in
the figures. For the weak example cell, \(\tau\) is undefined since
\(\widehat{\mathcal{L}} < 0\), and so the signifiance threshold is never
crossed in expectation.

Because \(\tau\) is a strictly decreasing function of \(\mathcal{L}\),
it ranks models identically to the heldout log-likelihood (and hence to
bits per spike). Thus, it is not a fundamentally new statistic, but a
more interpretable \emph{unit} for an existing one, answering the
question: ``how much data do I need to confirm that \(Q\) beats \(B\)?''
Quantities similar to the time to significance appear in prior
literature, notably Wald's sequential probability ratio test
(\citeproc{ref-wald1945}{Wald 1945}).

\section{Limitations}\label{limitations}

The material in this note comes with at least three caveats.

\emph{Variability across splits.} The trajectories in Figures 2--4 come
from ten independent train/test splits, so part of their spread reflects
variation in the \emph{fitted} model \(Q\) rather than variation in the
betting game itself. The game as formulated conditions on a fixed \(Q\);
reporting a distribution of \(\tau\) over splits is a practical hedge,
not a confidence interval for \(\tau\).

\emph{Independence and temporal structure.} The derivation assumes
heldout samples are i.i.d., which we enforced by shuffling time bins
before splitting. That discards the substantial autocorrelation present
in neural recordings. For models whose advantage over a baseline lies
precisely in capturing temporal dependence --- spike-history filters,
latent dynamics --- shuffling would misstate the evidence. The
martingale argument itself does not require independence, only that
contracts be priced under the conditional distribution implied by the
baseline given the past; extending the demonstration to sequentially
conditioned baselines a natural extension.

\emph{Composite nulls.} The null treated here, \(H_0 : P = B\), is
simple: it fixes a single baseline distribution. In practice \(B\) is
itself fit to training data, so a more honest null would be
composite---e.g., the family of all homogeneous Poisson processes.
Because \(B\) is estimated on training data independent of the test set,
the simple-null treatment is a reasonable approximation, but the
distinction could matter for small training sets and for more expressive
baselines. See Wasserman, Ramdas, and Balakrishnan
(\citeproc{ref-wasserman2020}{2020}), Grünwald, de Heide, and Koolen
(\citeproc{ref-grunwald2024}{2024}), and the monograph by Ramdas and
Wang (\citeproc{ref-ramdas2025}{2025}) for more rigorous treatment of
composite nulls.

\section{Code and data availability}\label{code-and-data-availability}

The analysis script that produced Figures 1--4, together with the
sources for this document, are available at
\url{https://github.com/neurostatsblog/neurostatsblog.github.io} under
\texttt{code/betting/} and \texttt{content/}. The recordings are the
publicly available mouse anterior-thalamus head-direction dataset of
Peyrache et al.~(2015), downloaded through the example-data facility of
\textbf{nemos}.

\section{Acknowledgements}\label{acknowledgements}

In addition to institutional support from New York University and the
Flatiron Institute (Simons Foundation), I am grateful to the McKnight
Foundation and the NIH BRAIN Initiative (1RF1MH133778) for financially
supporting my research program.

This note is an expanded version of a blog post by the author. A large
language model (Claude, Anthropic) was used to facilitate small portions
of the text and code. The framing and the mathematical arguments are the
author's own. The author has checked the full text, code
implementations, and references and is responsible for their accuracy.

\section{References}\label{references}

\phantomsection\label{refs}

\appendix

\section{Supplementary Note 1}\label{supplementary-note-1}

Here we sketch how the market price constraint
\eqref{eq:general-pricing-constraint} arises in more detail. For
simplicity, let's consider a case where the random outcomes \(X \sim P\)
take on one of \(n\) discrete values. That is, \(X \in \{1, \dots, n\}\)
almost surely. The same argument can be extended to continuous-valued
random variables with sufficient care.

By assuming there are only \(n\) discrete outcomes, then we can express
any potential contract function \(C(\cdot)\) as a finite linear
combination of elementary basis functions: \begin{equation}
C(x) = \sum_{i=1}^n r_i \delta_i(x)
\label{eq:contract-decomposition}
\end{equation} where \(r_1, \dots, r_n\) are scalar coefficients
denoting the return of outcome \(x = i\), \begin{equation}
r_i = C(i) ~,
\end{equation} and \(\delta_1, \dots, \delta_n\) are contracts that pay
off one unit of wealth if the outcome is \(x = i\), \[
\delta_i(x) = \begin{cases}
1 & x = i \\
0 & x \neq i
\end{cases} ~ .
\]

Recall that our goal is to show
\(\pi(C) = \mathbb{E} \left [ C(X) \right ]\), with the expectation
taken with respect to some appropriate choice of distribution \(B\). We
will show that this is true if the market is organized such that no
player can receive ``free money'' without taking any risk (in other
words, there are no arbitrage opportunities). We assume that any player
can either buy or sell contracts, and that fractional contracts are
supported in the market (e.g.~by sharing contracts with other players).

\textbf{Observation 1 -- The pricing function \(\pi\) must be linear.}
For any scalar \(a>0\), we must have \(\pi(a \cdot C) = a \pi(C)\). If
\(\pi(a \cdot C) > a \cdot \pi(C)\) a player would get free money by
selling the contract \(a \cdot C\) and simultaneously buying \(a\)
contracts of \(C\). For any outcome \(X \sim P\), the player neither
gains nor loses any wealth and they pocket
\(\pi(a \cdot C) - a \cdot \pi(C) > 0\) units of wealth. Conversely, if
\(\pi(a \cdot C) < a \cdot \pi(C)\) a player would get free money by
buying the contract \(a \cdot C\) and simultaneously selling \(a\)
contracts of \(C\). An arbitrage-free pricing scheme therefore must
satisfy \(\pi(a \cdot C) = a \pi(C)\).

Next, for any two contracts \(C_1\) and \(C_2\) we must have that
\(\pi(C_1 + C_2) = \pi(C_1) + \pi(C_2)\). The argument is quite similar
to above. If \(\pi(C_1 + C_2) > \pi(C_1) + \pi(C_2)\) then a player
would get free money by selling the contract
\(C_3(x) = C_1(x) + C_2(x)\) and simultaneously buying \(C_1\) and
\(C_2\). Since \(C_1(x) + C_2(x) - C_3(x) = 0\) for all \(x\), the
player neither gains nor loses any wealth based on the random outcome
and they pocket \(\pi(C_3) - \pi(C_1 + C_2) > 0\) units of wealth.
Conversely, if \(\pi(C_1 + C_2) < \pi(C_1) + \pi(C_2)\) then run the
same trade in reverse to get free money.

Taken together, we conclude that pricing must be linear. Applying this
to \eqref{eq:contract-decomposition} we can conclude that the price of
any contract can be written down as: \begin{equation}
\pi(C) = \sum_{i=1}^n r_i \pi(\delta_i)
\label{eq:contract-price-decomposition}
\end{equation} Thus, we can determine the price of any contract by
determining the prices of the elementary basis contracts
\(\delta_1, \dots, \delta_n\).

\textbf{Observation 2 -- the prices
\(\pi(\delta_1), \dots, \pi(\delta_n)\) are nonnegative and sum to one.}
First we prove nonnegativity. The random payout for each contract,
\(\delta_i(X)\) for \(X \sim P\), is greater than or equal to zero
almost surely. Thus, the price of each contract must be nonnegative---if
it were negative, it would imply that the player is \emph{paid} to
accept a contract that never results in a loss (i.e.~receive free
money).

Next we prove the normalization condition that
\(\sum_i \pi(\delta_i) = 1\). Note that we can construct a contract with
constant payoff \(C(X) = 1\), almost surely, by setting
\(r_1 = r_2 = \dots = r_n = 1\) in equation
\eqref{eq:contract-decomposition}. If the price of this contract were
less than one, a player would get free money by purchasing it. Likewise,
if the price were greater than one, a player would get free money by
selling it.

\textbf{Putting it together.} From observation 2, it is clear that the
prices \(\pi(\delta_1), \dots, \pi(\delta_n)\) define a probability
measure over outcomes \(x \in \{ 1, \dots, n \}\). Call this probability
measure \(B\). Then, recalling that \(r_i = C(i)\) denotes the return of
outcome \(x = i\) under the contract, we deduce from equation
\eqref{eq:contract-price-decomposition}: \[
\pi(C) = \sum_{i=1}^n r_i \pi(\delta_i) = \mathbb{E}_{X \sim B}  \left [ C(X) \right ]
\] confirming our claim that the price of a contract is given by the
expected payoff of the contract under an appropriate distribution \(B\).

\begin{center}\rule{0.5\linewidth}{0.5pt}\end{center}

\section{Supplementary Note 2}\label{supplementary-note-2}

We prove that the optimization problem stated in
\eqref{eq:kelly-criterion} is solved by the likelihood ratio
\(S^\star(x) = q(x)/b(x)\) given in
\eqref{eq:betting-function-equals-likelihood-ratio}.

The argument relies on a useful reparameterization. Let
\(r(x) = b(x) \, S(x)\) and note that \(r(x)\) is a probability density.
Indeed, \(r(x) > 0\) since both \(b\) and \(S\) are strictly positive,
and \[
\int r(x) \, dx \,=\, \int b(x) \, S(x) \, dx \,=\, \mathbb{E}_{X \sim B} [ S(X) ] \,=\, \pi(S) \, = \, 1
\] by the unit-price constraint \eqref{eq:unit-price-constraint}.
Conversely, since we assumed \(b(x) > 0\) everywhere on the support of
\(P\), any probability density \(r\) defines a feasible contract
function via \(S(x) = r(x)/b(x)\).

Now substitute \(S(x) = r(x)/b(x)\) into the objective in
\eqref{eq:kelly-criterion}, then add and subtract
\(\mathbb{E}_{X \sim Q} \log q(X)\): \begin{align}
\mathbb{E}_{X \sim Q} \log S(X)
  &\,=\, \mathbb{E}_{X \sim Q} \log \frac{r(X)}{b(X)} \\
  &\,=\, \mathbb{E}_{X \sim Q} \log \frac{q(X)}{b(X)} \,-\, \mathbb{E}_{X \sim Q} \log \frac{q(X)}{r(X)} \\
  &\,=\, D_{\mathrm{KL}}(Q \,\Vert\, B) \,-\, D_{\mathrm{KL}}(Q \,\Vert\, R) .
\end{align} The first term, \(D_{\mathrm{KL}}(Q \,\Vert\, B)\), does not
depend on the player's choice of contract. The second term,
\(D_{\mathrm{KL}}(Q \,\Vert\, R)\), is non-negative by Gibbs'
inequality, with equality if and only if \(R = Q\). To maximize the
objective, the player should therefore choose \(R = Q\), that is,
\(r(x) = q(x)\). Translating back to a contract function via \(S = r/b\)
yields \[
S^\star(x) \,=\, \frac{q(x)}{b(x)},
\] as claimed in \eqref{eq:betting-function-equals-likelihood-ratio}.
The maximum achievable value of the objective is
\(D_{\mathrm{KL}}(Q \,\Vert\, B)\) --- the exponential rate at which the
player anticipates their wealth will grow, recovering the
Kelly-criterion interpretation discussed in the main text.

\begin{center}\rule{0.5\linewidth}{0.5pt}\end{center}

\section{Supplementary Note 3}\label{supplementary-note-3}

Here we give an informal proof sketch of Ville's inequality. The
remarkable part of this result is that the inequality holds
\emph{uniformly over all time \(t\)}. So if we simulated a very large
number of wealth trajectories (i.e.~player \(Q\) gets to restart and
play the game many times), then only a small fraction, \(\alpha\), of
the trajectories would ever cross above \(1/\alpha\), \emph{even if each
individual trajectory were simulated infinitely long.}

Recall that we are given a discrete-time sequence \((M_t)_{t \geq 0}\),
satisfying \(\mathbb{E}[M_{t} \mid M_0, \dots, M_{t-1}] \leq M_{t-1}\)
and \(M_t \geq 0\) for all \(t\). A sequence with these two properties
is called a \emph{nonnegative supermartingale}.

Let \(c\) denote the first time that the trajectory crosses above a
threshold \(\lambda > 0\). That is, \(c\) is the smallest natural number
such that \(M_c \geq \lambda\). If the trajectory \emph{never} crosses
above \(\lambda\), we take this as meaning \(c = \infty\). Our goal is
to prove \begin{equation}
\textrm{Pr}(c~\text{is finite}) \leq \frac{\mathbb{E} [M_0]}{\lambda}
\label{eq:ville-simplified}
\end{equation} which is more-or-less identical to the result we want. In
the main post, we chose the threshold to be \(\lambda = 1/\alpha\).

The trick is to define a new \emph{nonnegative supermartingale}
\((Z_t)_{t \geq 0}\) as \begin{equation}
Z_t = 
\begin{cases}
M_t & \text{if $t < c$} \\
\lambda & \text{if $t \geq c$} \\
\end{cases}
\end{equation} Note that if \(c = \infty\), then \(M_t\) never crosses
above the threshold and \(Z_t\) is just a copy of \(M_t\) for all \(t\).
It is easy to show that if \((M_t)_{t \geq 0}\) is a nonnegative
supermartingale, then so is \((Z_t)_{t \geq 0}\) (exercise to the
reader).

Now for any fixed value of \(t\), we can apply Markov's inequality to
conclude that \begin{equation}
\textrm{Pr}(c \leq t) = \textrm{Pr}(Z_t \geq \lambda) \leq \frac{\mathbb{E} [Z_t]}{\lambda} \leq \frac{\mathbb{E} [Z_0]}{\lambda}
\end{equation} The final inequality follows from the fact that
\((Z_t)_{t \geq 0}\) is a nonnegative supermartingale, which implies
\(\mathbb{E} [ Z_t ] \leq \mathbb{E} [ Z_0 ]\).

Further, \(\mathbb{E} [Z_0] \leq \mathbb{E} [M_0]\), since \(Z_t\) is a
clipped version of \(M_t\). Thus, we have: \begin{equation}
\textrm{Pr}(c \leq t) \leq \frac{\mathbb{E} [M_0]}{\lambda}.
\end{equation} Notice that the right hand side is not a function of
\(t\). There is a way to rigorously take the limit of
\(t \rightarrow \infty\) on both sides of the inequality to yield our
desired result, equation \eqref{eq:ville-simplified}.

\end{document}